\documentclass[preprint,11pt]{article}

\usepackage{bm}
\usepackage{epsfig}
\usepackage{color}
\usepackage{amssymb}
\usepackage{amsmath}

\newcommand {\lab}[1]{\label{eq:#1}}

\newcommand {\be}[1]{\begin{equation}{\lab{#1}}}
\newcommand {\ee}{\end{equation}}
\newcommand {\bea}{\begin{eqnarray}}
\newcommand {\eea}{\end{eqnarray}}

\begin{document}

\title{The effect of long--range interactions on the dynamics and statistics of 1D Hamiltonian lattices with on--site potential}

\author{
\textbf{ H. Christodoulidi$^{1}$, T. Bountis$^{2}$,  L. Drossos$^{3}$}\\
$^{1}$Research Center for Astronomy and Applied Mathematics  
\\ Academy of Athens, Athens, Greece\\ 
University of Patras, GR-26500 Patras, Greece \\
$^2$Department of Mathematics, Nazarbayev University, \\
Astana, Republic of Kazakhstan\\
$^3$High Performance Computing Systems and Distance Learning Lab,\\
Technological Educational Institute of Western Greece, Greece }

 \maketitle

\abstract{
We examine the role of long--range interactions on the dynamical and statistical properties of two 1D lattices with on--site potentials that are known to support discrete breathers: the Klein--Gordon (KG) lattice which includes linear dispersion and the Gorbach--Flach (GF) lattice, which shares the same on--site potential but its dispersion is purely nonlinear. 
In both models under the implementation of long--range interactions (LRI) we find that 
single--site excitations lead to special low--dimensional solutions,
which are well described by the undamped Duffing oscillator.
For random initial conditions we observe that the maximal Lyapunov exponent $\lambda $
scales as $N^{-0.12}$ in the KG model and as $N^{-0.27}$ in the GF with LRI, suggesting in that case an approach  to integrable behavior towards the thermodynamic limit. Furthermore, under LRI, their non-Gaussian momentum distributions are distinctly different from those of the FPU model.}

\section{Introduction}
\label{intro}
 In recent years there has been great interest in many--particle systems whose components interact via long--range forces 
\cite{Anteneodo,latora,CirtoAssisTsallis2013,EPL,JSTAT,Ruffo13,Ruffo14,Ruffo15,Gupta,Iubini,Bagchi,Bagchi0,Bagchi1,Bagchi2}. 
Such models exhibit various forms of organization, such as synchronization, collective chaos, long--living quasi--stationary states and often, even a global decrease of chaos as the number of particles increases with constant energy per particle $\varepsilon =E/N$. An extensively studied Hamiltonian model of this type with long--range forces is the so--called Mean Field Hamiltonian (HMF),
which is composed by $N$ identical particles lying on a ring \cite{Anteneodo,latora,CirtoAssisTsallis2013}. In this globally coupled system each particle interacts equally with all the others, independently of the topological distance among the sites. An important result of these studies \cite{Anteneodo,latora} is that the maximal Lyapunov exponent scales with the degrees of freedom like $\lambda \sim N^{-1/3}$, suggesting a non--ergodic behavior towards the thermodynamic limit.
 
More recently, the authors of the present paper, together with C. Tsallis, studied a long--range interaction generalisation of the one-dimensional (1D) Fermi--Pasta--Ulam (FPU) $\beta$--model \cite{feretal1955}, by introducing a quartic interaction coupling constant that decays as $1/r^\alpha$ ($\alpha \ge 0$), with $\alpha \to\infty$ corresponding to the nearest--neighbor FPU model \cite{EPL}. Through molecular dynamics, it was shown that: (i) For $\alpha \geq 1$ the maximal Lyapunov exponent is finite and positive for increasing number of particles $N$, whereas, for $0 \le \alpha <1$, it asymptotically decreases as a power law in $N$. (ii) The distribution of time-averaged velocities is Gaussian for $\alpha$ large enough, whereas it becomes a $q>1$-Gaussian, when $0\leq \alpha<1$, thus leading for $\alpha$ small enough to a crossover from 
$q$--statistics \cite{Tsallis1988,GellMannTsallis2004,Tsallis2009,Tsallis2014} to Boltzmann--Gibbs (BG) thermostatistics as time increases. 

In a subsequent paper, the same FPU model was revisited introducing long--range interactions (LRI) in both the quadratic and quartic parts of the potential, via two independent exponents $\alpha_1$ and $\alpha_2$. It was demonstrated that weak chaos, in the sense of decreasing Lyapunov exponents and $q$--Gaussian  probability density functions (pdfs), occurs only when long--range interactions are included in the quartic part \cite{JSTAT}, while, under certain conditions, these pdfs are found to persist as $N\rightarrow \infty$. On the other hand, when LRI are imposed only on the quadratic part, strong chaos and purely Gaussian pdfs are always obtained. In fact, in the extremal case where phonons are completely absent \cite{Bagchi}, the FPU system under LRI exhibits the same `non-ergodic' behavior, with $\lambda \sim N^{-1/3}$ towards the thermodynamic limit, as in the HMF model. 

It is important to recall, of course, that the HMF and FPU models are characterized by translational invariance. Thus, in the present paper, we investigate whether any of the above phenomena also occur in Hamiltonian 1D lattices possessing on--site potentials. As is well--known, such systems typically exhibit localized oscillations called discrete breathers. It would be challenging to extend previous studies of the impact of LRI on the dynamics and statistics of such models. To this end, we focus here on two 1D Hamiltonians with on--site potentials, namely, the Klein--Gordon (KG) and Gorbach--Flach (GF) models \cite{GFcite,Maniadis}. Inspired by our previous studies, we add to the linear dispersion of these models quartic interactions, and proceed to investigate their effect on the dynamics, by computing Lyapunov exponents, and statistics, by analyzing their momentum distributions for increasingly $N$ values. 
 
The present paper is organized as follows: In Section \ref{1} we describe the two models, KG and GF, and study the energy spreading caused by the initial excitation of a single particle,
when  nearest neighbor-interactions apply to both models (Subsection \ref{2.1}) and when the longest--range 
interactions are applied with $\alpha =0$ (Subsection \ref{2.2}). 
We then proceed to more general initial conditions and investigate the values of the maximal Lyapunov exponent
under short and long--range interactions in Section \ref{gc}. In particular, we study 
the behavior of the maximal Lyapunov exponent in all cases where the energy is gradually increased (Subsection \ref{gc1})
and separately, when the number of particles increases (Subsection \ref{gc2}) at fixed energy per particle.  
Finally, in Section \ref{md} we examine the corresponding momentum distributions of these systems characterizing their statistical properties, and discuss our findings in Section \ref{concl}.

\section{The KG and GF Hamiltonians with on site potential \label{1}}

We shall consider two types of 1D Hamiltonian lattices which share the same on--site potential but differ in the degree of their dispersive terms. The first one is the classic Klein--Gordon (KG) lattice, with quadratic and quartic on site potential and {\it linear} dispersion terms, described by the Hamiltonian:
\begin{equation}\label{KG}
{H}^{KG}(p,x)= \sum_{n}  \frac{1}{2} p_n^2 + \frac{1}{2} x_n^2 + \frac{1}{4} x_n^4 +  \frac{C }{2} (x_{n+1}-x_n)^2 .
\end{equation}
The second is the Gorbach--Flach  model (GF) \cite{GFcite} consisting of $N$ coupled oscillators 
with the same KG on--site potential, but with {\it only quartic} interactions among its nearest--neighbor sites, 
\begin{equation}\label{GF}
{H}^{GF}(p,x)= \sum_{n} \frac{1}{2} p_n^2 + \frac{1}{2} x_n^2 + \frac{1}{4} x_n^4 +  \frac{C }{4} (x_{n+1}-x_n)^4 ,
\end{equation}
and is hence free from linear dispersion. In both systems $p_n$ and $x _n$ represent the canonical conjugate pairs of momenta and positions respectively, satisfying periodic boundary conditions $x_0=x_N, x_{N+1}=x_1$. In short, the two Hamiltonians  (\ref{KG}) and (\ref{GF}) have the form:
\begin{equation}\label{ham}
{H} (p,x)= \sum_{n}  \frac{1}{2} p_n^2 +V(x_n) + W(x_{n+1}-x_n)=E
\end{equation}
where $E$ is the total energy. By $V(x)$ we denote the so--called `hard' on--site potential $\frac{1}{2} x^2 + \frac{1 }{4} x^4$, present in both systems, while $W(x)$ denotes the potential functions $W^L(x)=\frac{C }{2} x^2$ and $W^N(x)= \frac{C }{4} x^4$  describing linear and nonlinear coupling in KG and GF, respectively. 

\begin{figure*}
\centering
\resizebox{0.75\columnwidth}{!}{
\includegraphics{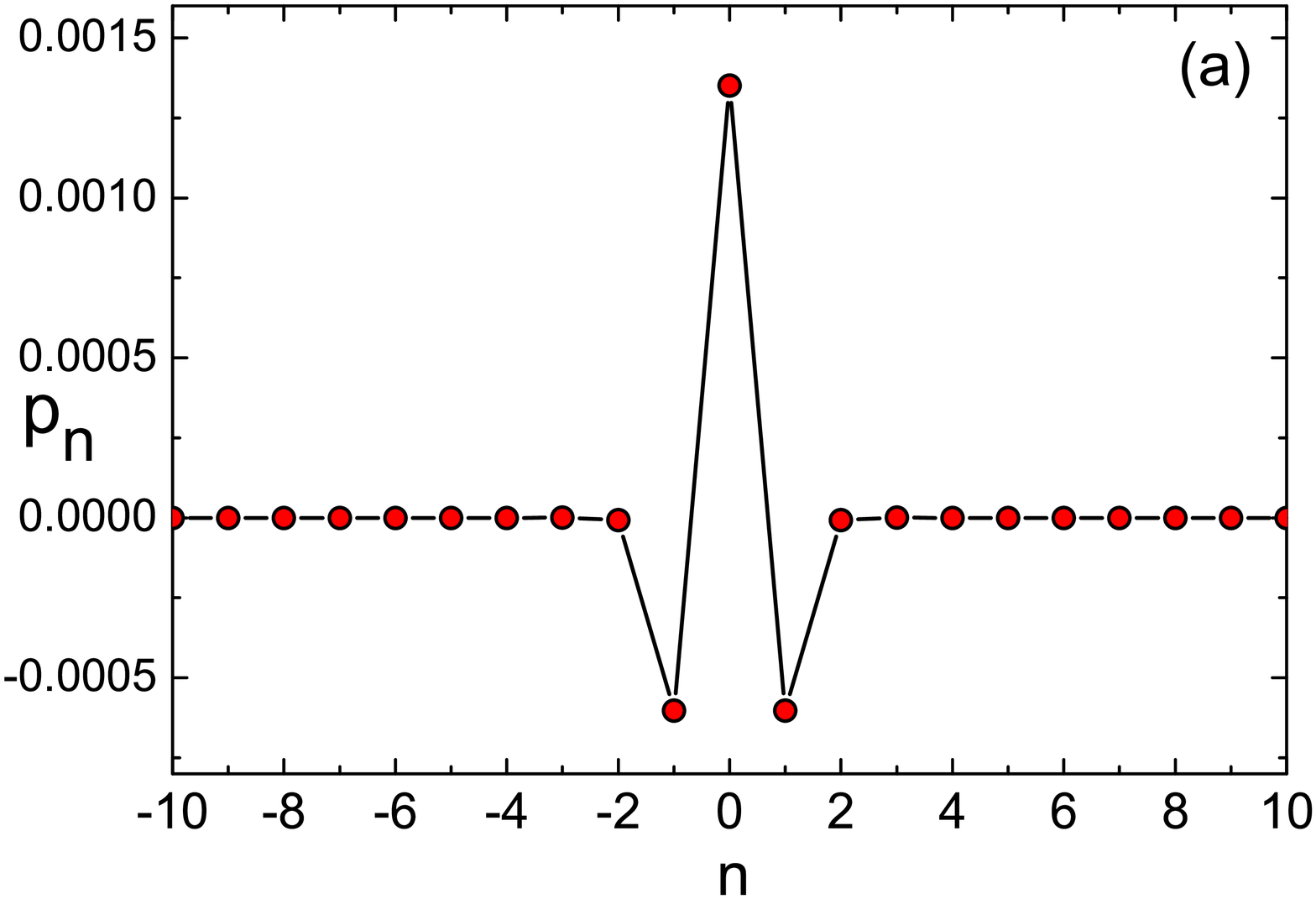} 
\includegraphics{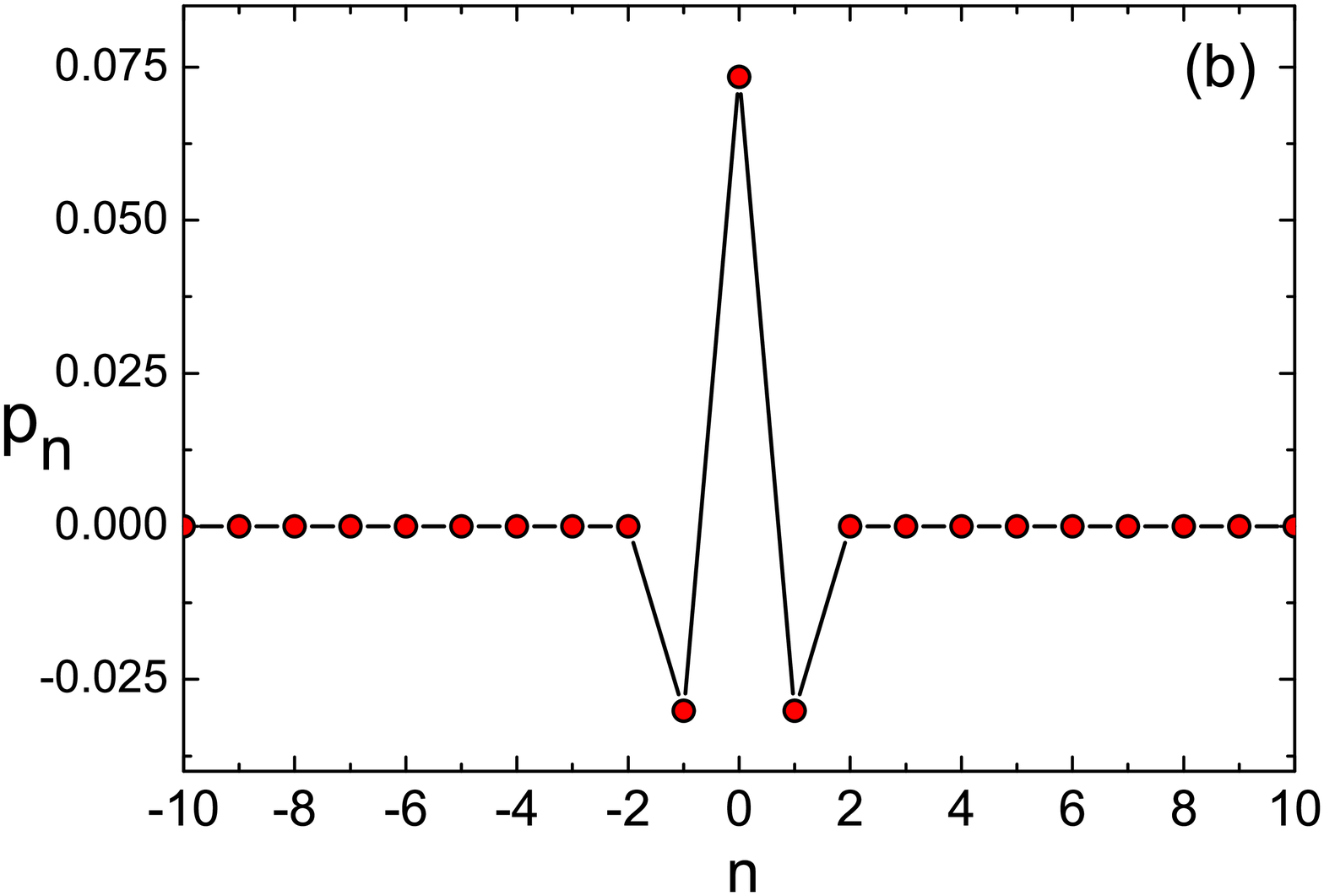}}
\caption{ A breather--like profile obtained a short time after a single--site excitation ($n=0$)  in 
(a) the Klein--Gordon model and (b) the Gorbach--Flach model, for $N=256$ particles.
\label{lattice}}
\end{figure*}

As is well known, these systems can support localized periodic oscillations called {\it discrete breathers}, 
centered about any one of their sites \cite{MA94,F95a,BBJ,BBV02}. The simplest ones among them are easy to find, by judiciously exciting a single site, provided their frequency lies outside the linear frequency spectrum of the lattice. If they are stable, their approximate form will be preserved in time, while if they are unstable they will eventually collapse and share their energy with all particles of the lattice. In Fig.~\ref{lattice} we display two examples of such breathers for the KG and GF models studied in this paper.

\subsection{Energy spreading after single--site excitations \label{2.1}}

\begin{figure*}
\centering
\resizebox{0.75\columnwidth}{!}{
\includegraphics{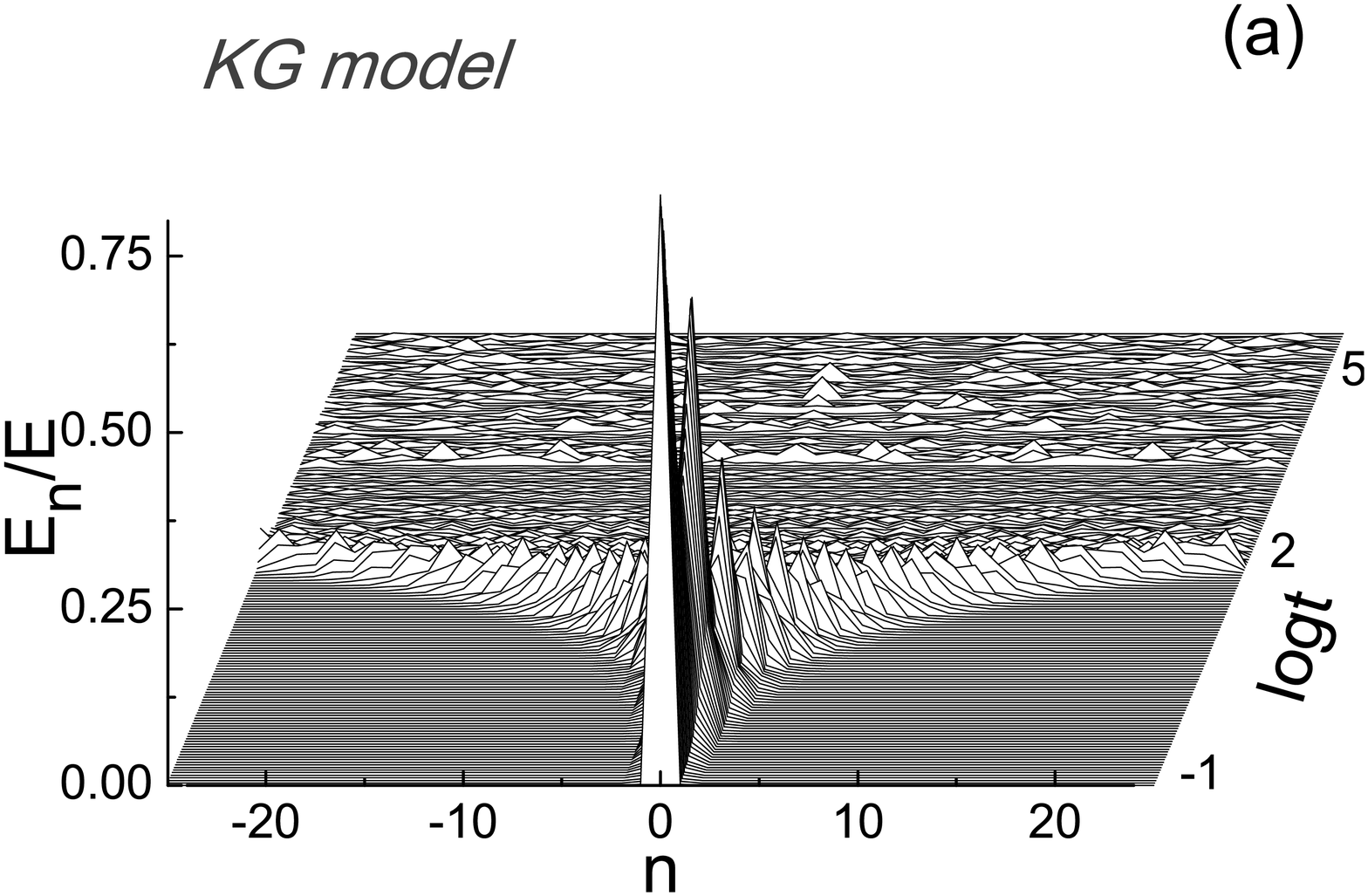} 
\includegraphics{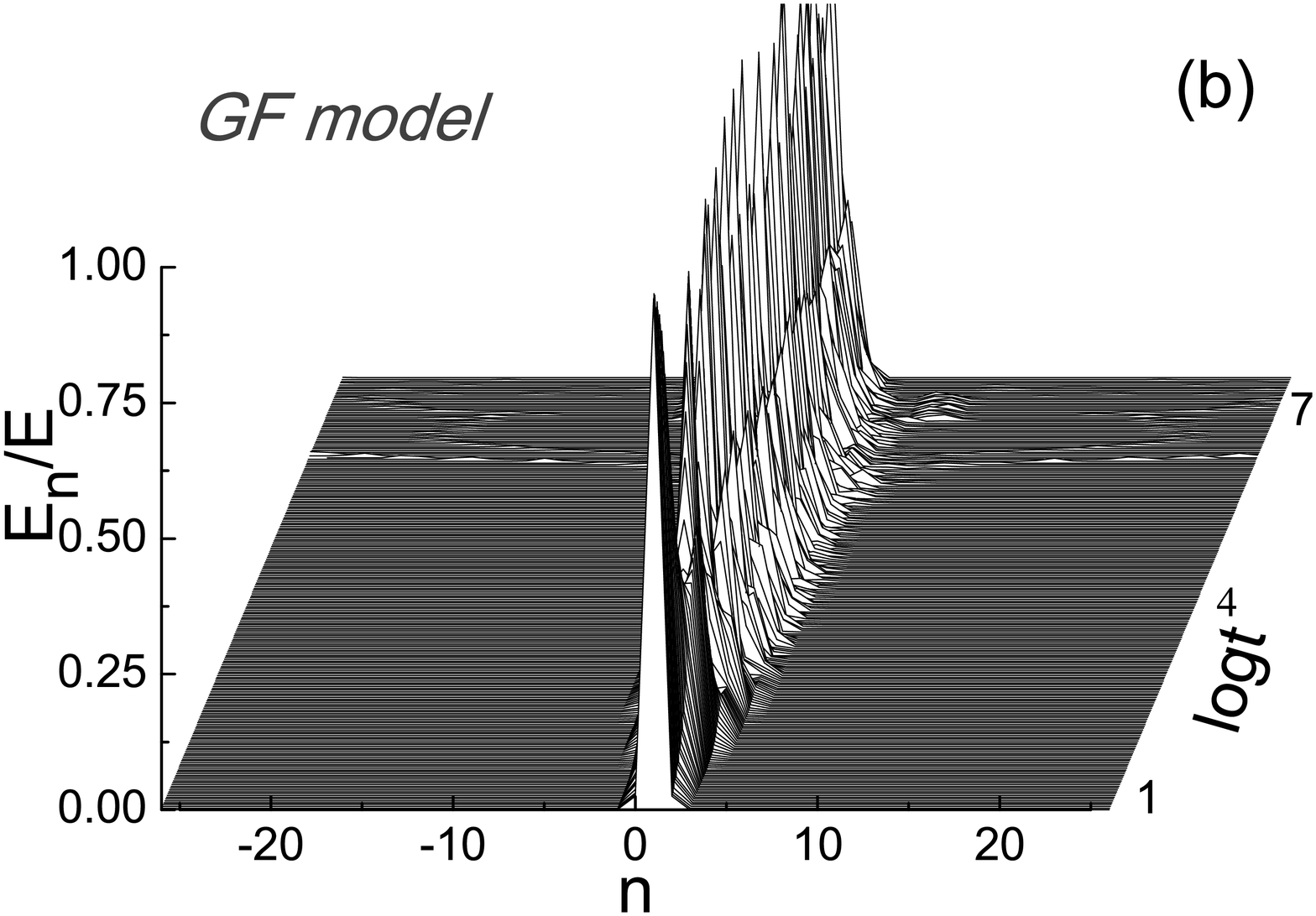}} 
\resizebox{0.75\columnwidth}{!}{
\includegraphics{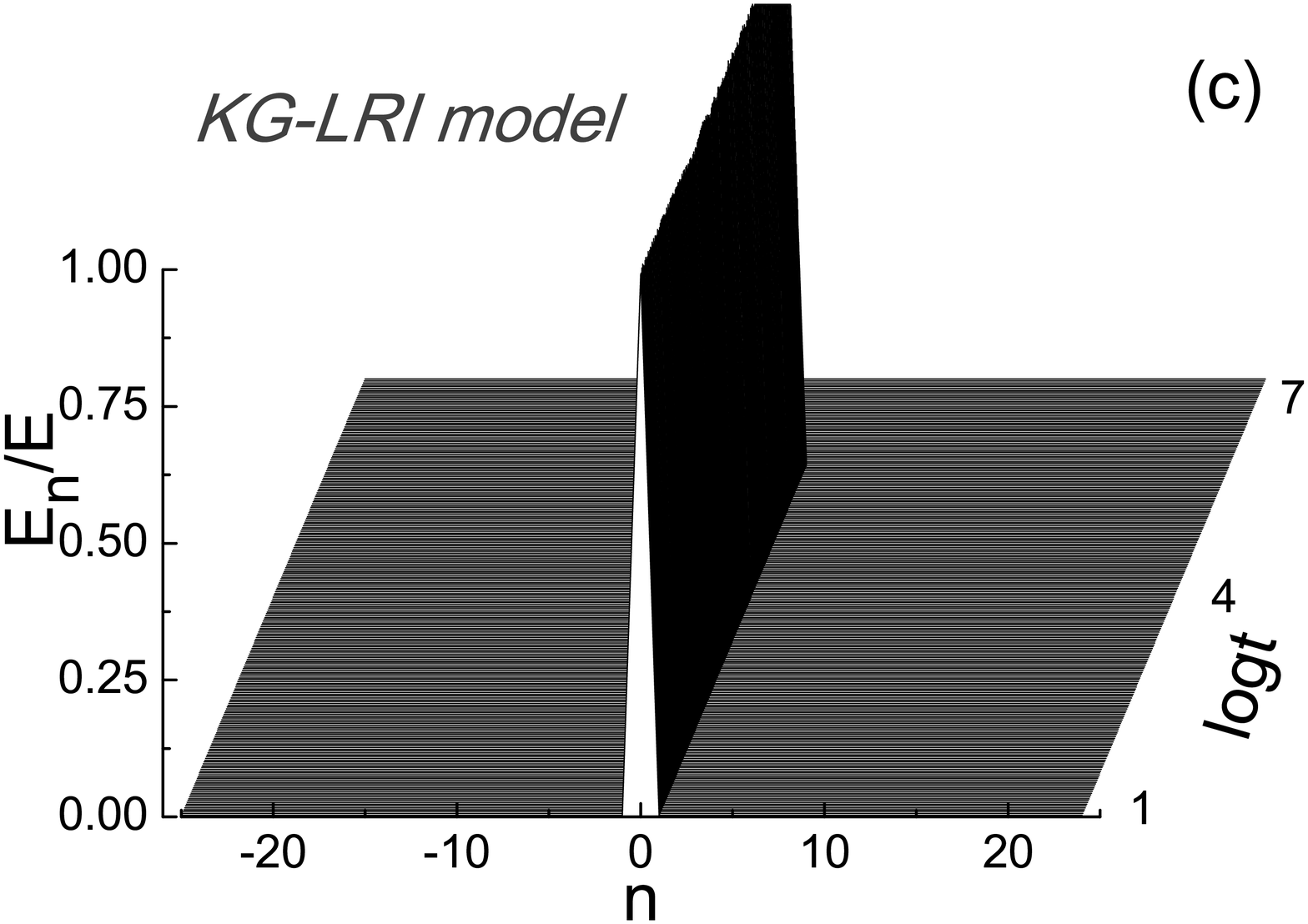} 
\includegraphics{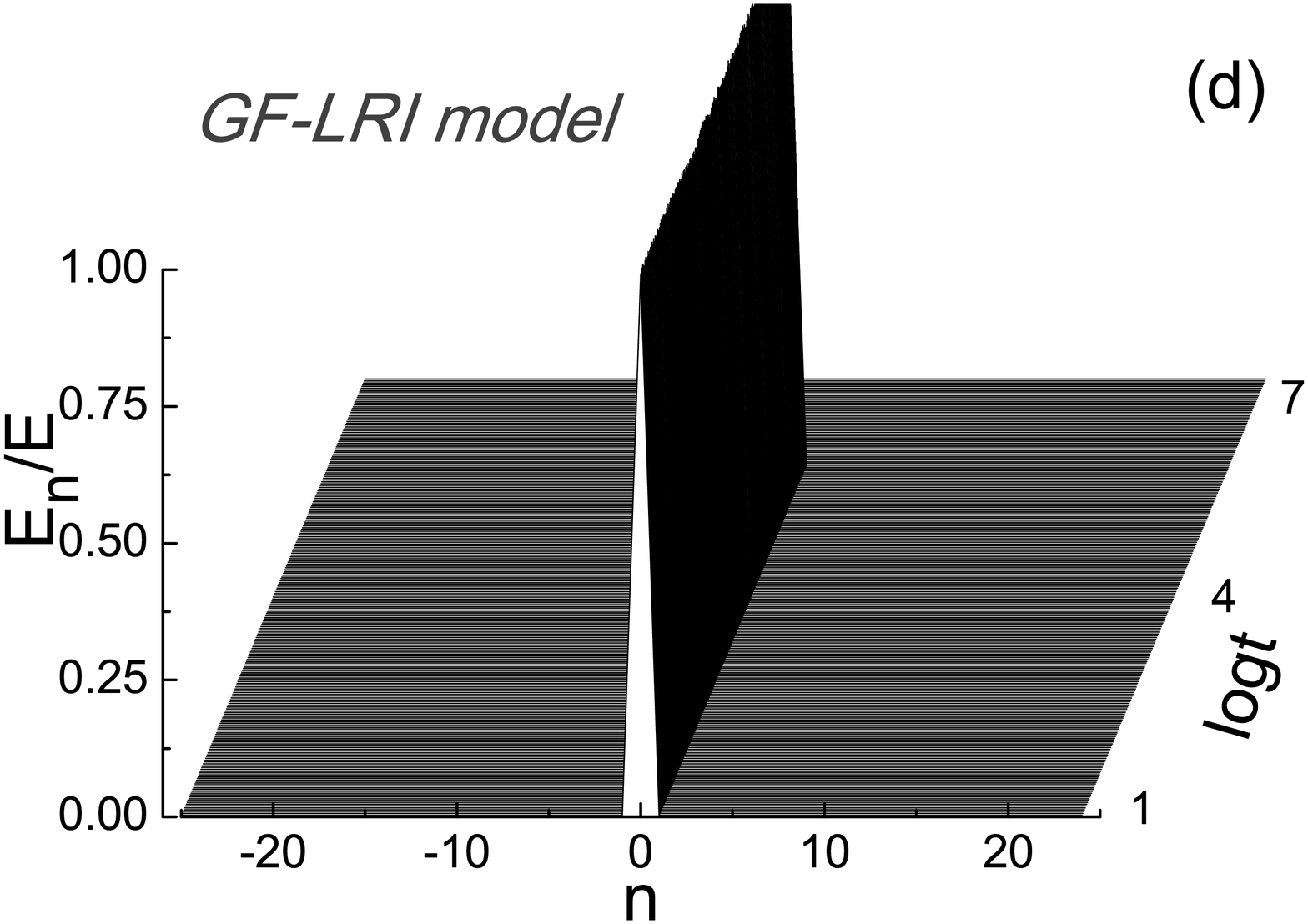}}
\caption{ The temporal evolution of the normalized on--site energies in (a) the classic Klein--Gordon  
and (b) the Gorbach--Flach model,  (c) the Klein--Gordon with LRI and 
(d) the Gorbach--Flach model with LRI. 
All parameter values are the same, i.e. $N=256$, $C=1$ and $p_0(0)=0.2$ .
\label{vfg}  }
\end{figure*}

Let us now focus on the effect of linear and nonlinear dispersion in the dynamics 
of the KG and GF lattices separately, following a single--site excitation of the 
central particle by a momentum shift of $p_{0}(0)=A$. Due to the absence of linear 
dispersion in the latter model, it is clear from Fig.~\ref{vfg} that the GF breather 
is significantly more robust. 

Energy spreading can be clearly monitored by the temporal evolution of the on--site energies $E_n$, given by
$$E_n=\frac{1}{2} p_n^2 + \frac{1}{2} x_n^2 + \frac{1}{4} x_n^4~~~. $$
Particle interactions have not been included, as usually is found in the literature \cite{Flach2004}
for instructive reasons, which ease the comparison between models with different potentials. 
In both models, shown in Fig.\ref{vfg}(a) and (b), we consider the same parameters, 
namely $N=256$ particles with coupling constant $C =1$ and and an initial momentum shift $p_0(0)=0.2$. 

Comparing the linear and the nonlinear dispersion in the first two panels of Fig.\ref{vfg}, 
we observe that the KG model quickly experiences a fast energy transfer caused by
the existence of a non-trivial harmonic spectrum, in contrast to 
the GF model where localization persists for very long times and energy propagation is much less pronounced. 
The near absence of diffusion in the latter case can be attributed to the lack of a linear frequency band and 
hence the absence of breather--phonon resonances.

\subsection{Energy spreading under long--range interactions \label{2.2}}
Let us note now that our KG and GF 1D Hamiltonians, as they appear in (\ref{KG}) and (\ref{GF}), 
involve nearest--neighbor interactions. It would, therefore, be interesting to ask how do their dynamical 
and statistical properties change if we extend the range of interactions and introduce forces that 
decay with distance according to $1/r^{\alpha }$. This implies an appropriate modification of the potential function $W(x)$.
The LRI--Hamiltonian thus becomes:
\begin{equation}\label{man2}
{H}(p,x)= \sum_{n}  \big( \frac{1}{2} p_n^2 + V(x_n) \big) + \frac{1}{2\tilde{N}}\sum_{n,m} \frac{W(x_{m}-x_n)}{r_{n,m}^{\alpha }}  ,
\end{equation}
where $r_{n,m}=min \{ \mid n-m\mid, N- \mid n-m \mid\} $ defines the minimum topological distance between 
the particles $n$ and $m$ and $\tilde{N}=1/N\sum_{n,m} 1/ {r_{n,m}^{\alpha}}$ provides a rescaling factor that maintains extensivity in the Hamiltonian (\ref{man2}), i.e. that all its parts grow $\propto N$.

In the remainder of the paper we focus only on the particular $\alpha =0$ case,  for which, the LRI--Hamiltonian (\ref{man2}) simplifies to:
\begin{equation}\label{man3}
{H}(p,x)= \sum_{n} \big( \frac{1}{2} p_n^2 + V(x_n) \big) + \frac{1}{2\tilde{N}}\sum_{n,m} W(x_m-x_n)   ,
\end{equation}
with $\tilde{N}=N-1$. By {\it KG--LRI and GF--LRI} we refer to the Klein--Gordon and the Gorbach--Flach models 
with long--range interaction, as described by the Hamiltonian (\ref{man3}) for appropriate $W(x)$ potential functions. 

\subsubsection{Single--site excitations for $\alpha =0$}
As has been convincingly shown in \cite{Flachlri}, that long-range interactions destroy the 
{\it strong localization profile} of discrete breathers, in the sense that, the exponential tails 
of the breather become algebraic. On the other hand, in the extreme LRI case of $\alpha =0$ adopted here, we find that the implementation of LRI is particularly beneficial, leading to the {\it persistence} of energy localization for very long times, as we explain below.

 
One interesting observation concerning breather generation and collapse can be made by imposing  
single--site excitations of appropriate magnitude to our new lattices with Hamiltonians of the form (\ref{man3}). 
For example, in Fig.\ref{vfg}(c) and (d) we repeat the calculations of Fig.\ref{vfg}(a) and (b) 
for the long--range cases in KG and GF keeping all parameters the same and we find that, 
the single--site excitation $p_0(0)=A$ results in a remarkably stable breather--like solution, 
which shows no sign of collapse even after very long times.

What in fact happens in Fig.\ref{vfg}(c), (d) is {\it not} exponential localization of a discrete breather, as we know it from the nearest neighbor cases of Fig.\ref{vfg}(a), (b). What we see in Fig.\ref{vfg}(c), (d) is the large amplitude oscillations of a single excited particle, say $n=0$, accompanied by much smaller oscillations of all other 
particles, moving with {\it equal} amplitudes.

Indeed, the kind of excitations observed in Fig.\ref{vfg}(c), (d) represent small oscillations about a simple but fundamental periodic solution of the equations of motion:
\begin{eqnarray}\label{a=0}
\ddot{x}_n &=& -x_n - x_n^3 + \frac{C }{\tilde{N}}(x_0 - x_n)^p +\frac{C }{\tilde{N}} \sum_{m \ne 0} (x_m - x_n)^p,
\end{eqnarray}
where $p=1$ for KG and $p=3$ for GF. In this solution, all small oscillations are equal, 
i.e. $x_m(t)=x(t)$ for all $m\ne 0$ and our system of equations of motion (\ref{a=0}) collapses to 
a two-degree of freedom system:
\begin{eqnarray}\label{sigma}
\ddot{x}_0 &=& -x_0 - x_0^3 + C (x - x_0)^p \nonumber\\
\ddot{x} &=& -x - x^3 - \frac{C }{ \tilde{N}} (x - x_0)^p~.
\end{eqnarray}
It would be straightforward to study its stability numerically by computing the associated $4\times 4$ monodromy matrix, or better yet display pictorially the orbits on a 2--dimensional surface  of section plot $x_0,\dot{x_0}$. We have carried out such plots and found that for the parameter values of interest there are large islands of stability about this orbit which guarantee its nonlinear stability even for large perturbations. 

It follows from the system (\ref{sigma}) that the oscillations $x(t)$  are of the order ${\cal O} (1/N)$, 
and hence, in the thermodynamic limit $N\rightarrow \infty $,  the network will be governed by the (one--degree of freedom) Duffing equation: 
\begin{eqnarray}\label{duff}
\ddot{x}_0 + \omega_0^2  x_0 + b x_0^3=0~.
\end{eqnarray} 
The solutions of (\ref{duff}) obtained by setting $p_0(0)=A<<1$ can be satisfactorily approximated by a single mode $x_0(t)=A\sin( \omega t)$, if we neglect higher harmonic terms 
of the form $\sin( 3\omega t)$ with coefficients of the order ${\cal O}(A^3)$. We thus obtain a first order correction of the frequency $$\omega \approx \omega _0 +\frac{3bA^2}{8\omega _0}$$
and by the energy conservation law we can get a rough estimate of the amplitude $A=\sqrt{2E}/\omega $.

\begin{figure*}
\centering
\resizebox{0.75\columnwidth}{!}{\includegraphics{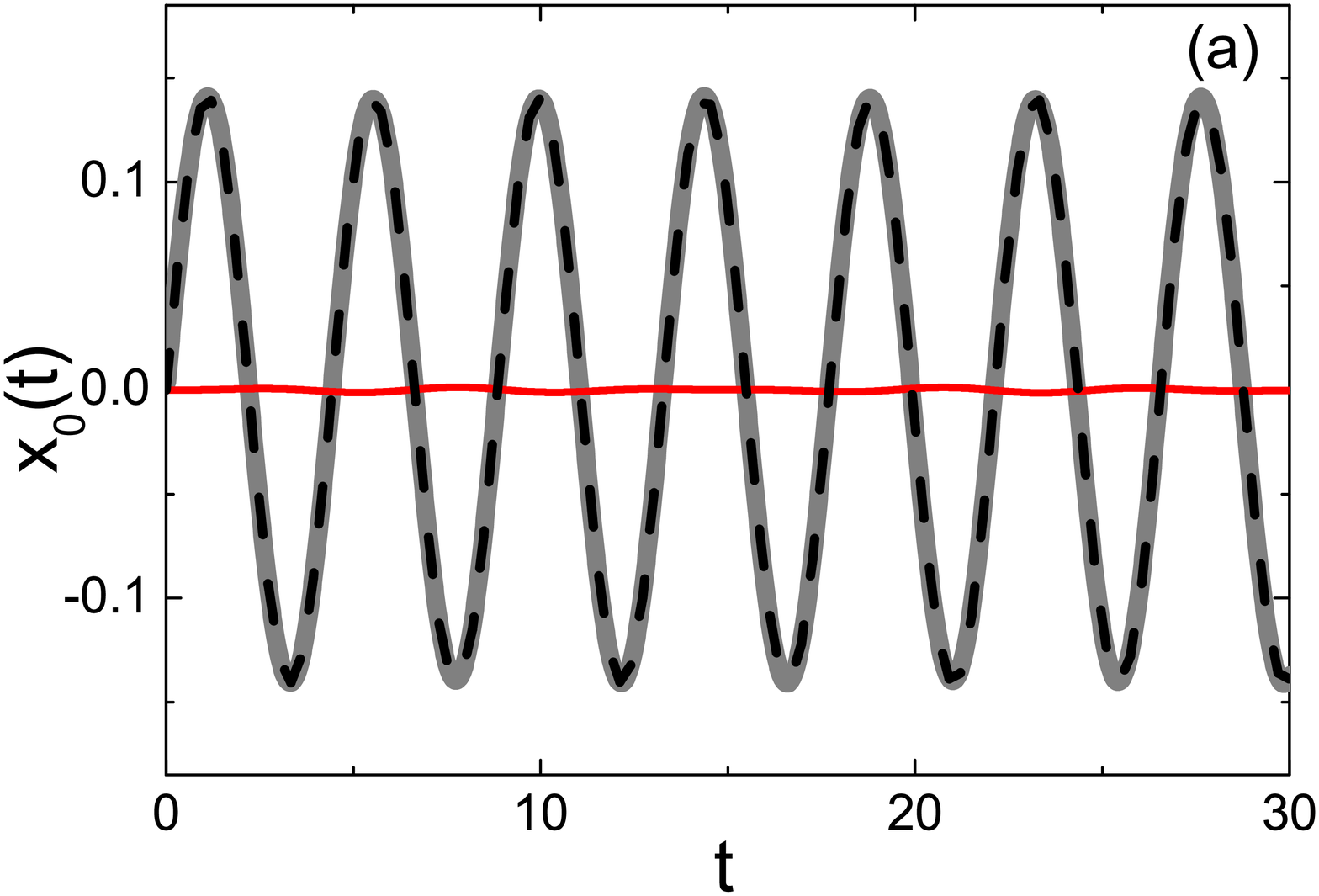}
\includegraphics{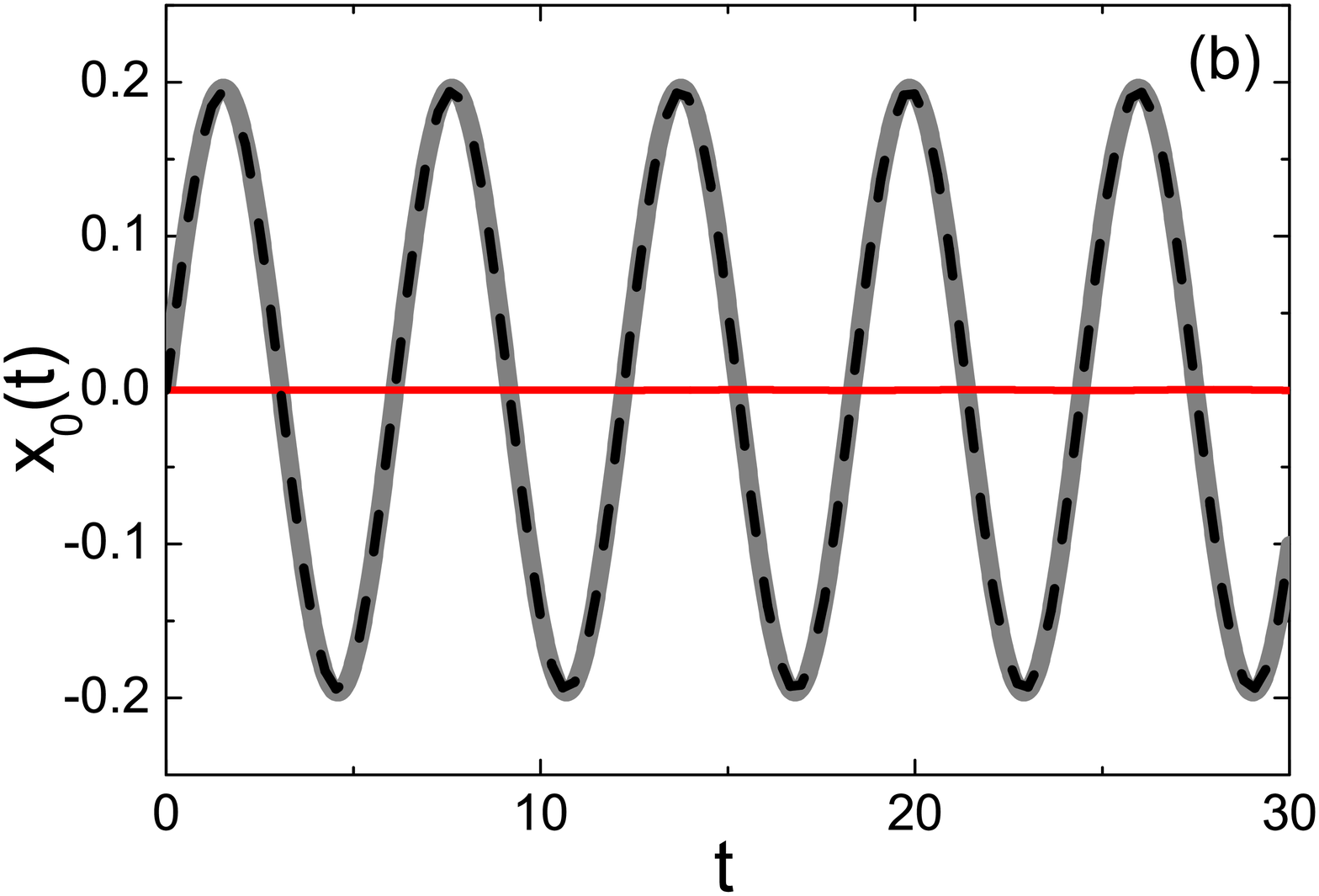}}
\caption{ The solution for the systems (a) KG--LRI and (b) GF--LRI of Fig.\ref{vfg}(c) and (d), respectively. 
In both panels the gray line is $x_0(t)$, the dashed black line is the estimated periodic solution $A\sin( \omega t)$ and with red are the rest
particles $x_n(t)\equiv x(t)$, for all $n\ne 0$. \label{oscillations}  }
\end{figure*}

In the two examples studied in Fig.\ref{vfg}(c), (d) our parameters have the values $(\omega_0^2,b)=(2,1)$ for KG--LRI 
model and $(\omega_0^2,b)=(1,2)$ for GF--LRI, and the above simple derivation yields the estimates $(A,\omega ) = (0.1409,1.4195)$ and  $(A,\omega ) = (0.194483,$ $1.02837)$, respectively,  which are excellent fits for the numerical solutions displayed in Fig.\ref{oscillations}.
   
The above theoretical framework explains the persistence of energy localization in a single particle observed in Fig.\ref{vfg}(c), (d), under uniform all--to--all LRI. Furthermore, there is no sign of collapse of the numerical solution, even after very long integration times, which implies that the exact periodic solution we found in its vicinity that causes the localization, is stable under small perturbations.

\section{Dynamics under short and long--range interactions with $\alpha=0$  \label{gc}} 

As explained in the Introduction, in the case of translationally invariant Hamiltonians like the FPU and HMF models, LRI has important implications regarding their dynamical and statistical behavior towards the thermodynamic limit. 
In particular, for $0\leq \alpha<1$, maximal Lyapunov exponents $\lambda$ exhibit saturation phenomena 
as the energy per particle $\varepsilon$ increases, while they tend to decrease to zero with increasing $N$. 

On the other hand, LRI are also associated with a change in the thermostatistics, since, for $0\leq \alpha<1$, momentum probability distributions (pdfs) deviate from Gaussians and approach $q>1$--Gaussians, characterized by long tails and algebraic power laws \cite{EPL,JSTAT}. We have attributed to this kind of regularization the term ``weak chaos'', as opposed to `strong chaos' 
associated with non--decreasing $\lambda$ and Gaussian pdfs related to exponential decay of correlations.

Let us now investigate whether similar effects occur in the present models when we estimate maximal Lyapunov exponents as functions of $\varepsilon, N$ and momentum pdfs as they evolve to a stationary shape, starting as always from initial momenta drawn randomly from a uniform distribution.

\subsection{Small and large energy limits \label{gc1}} 
\begin{figure*}
\centering
\resizebox{0.75\columnwidth}{!}{
\includegraphics{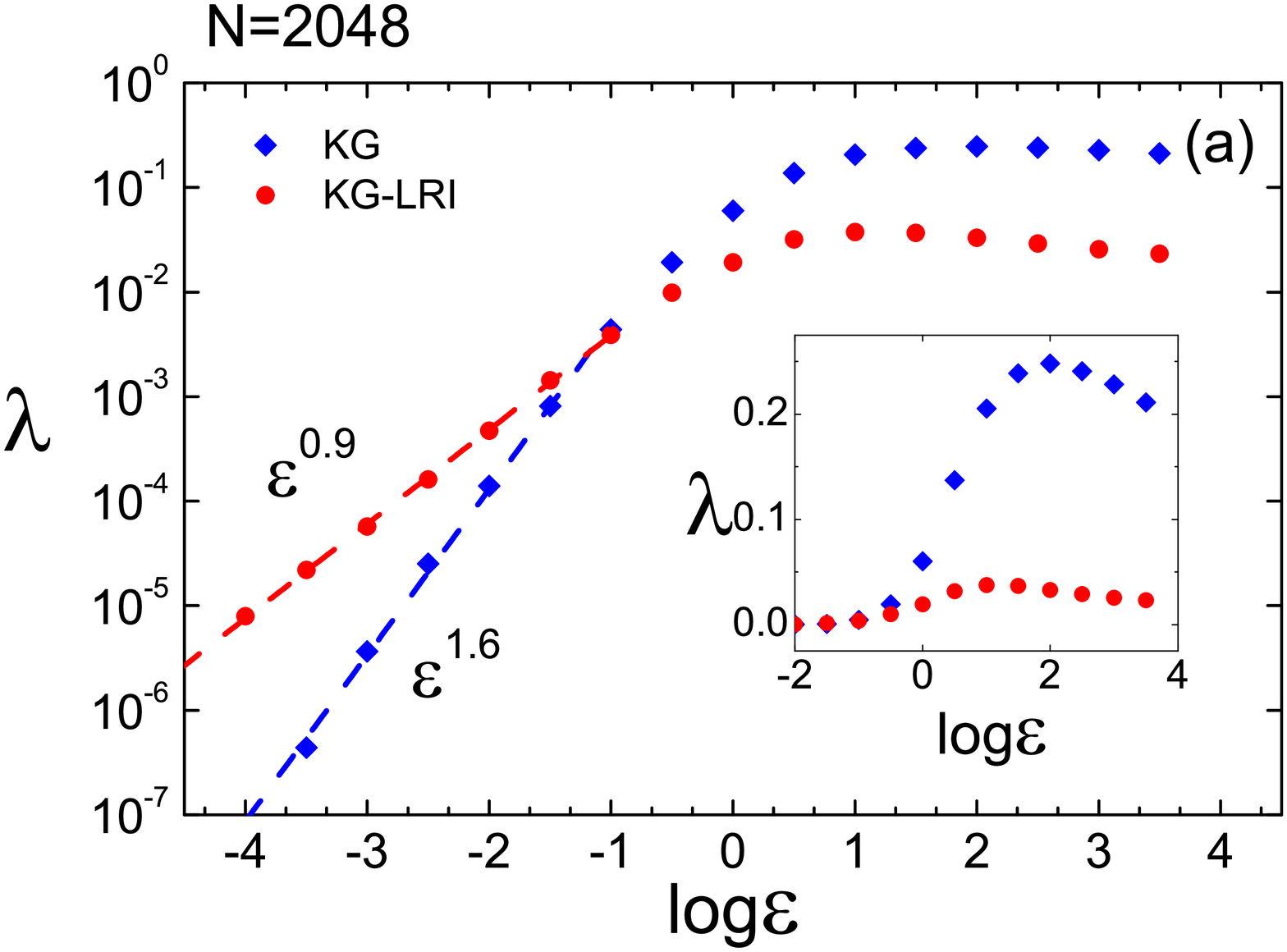} 
\includegraphics{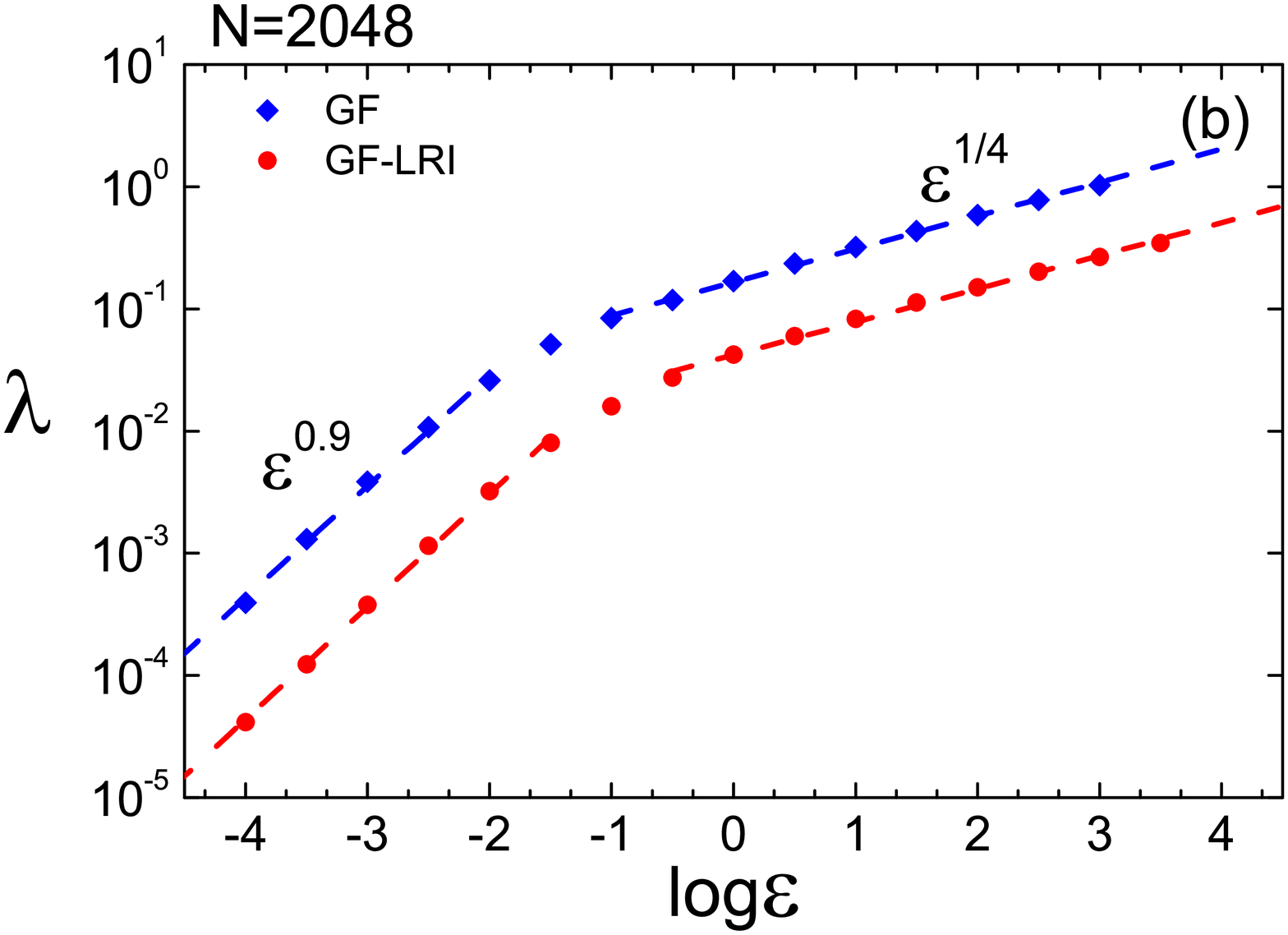} } 
\caption{ The maximal Lyapunov exponent $\lambda $ for increasing energy values of $\varepsilon $ 
and fixed number of particles $N=2048$. (a) 
The classic KG model (blue diamonds) and the KG model with LRI (red circles).
Inset: a closer look to the $\lambda $ values at high energies.
(b) The  GF model (blue diamonds) and the GF model with LRI (red circles).
\label{KGlyap}  }
\end{figure*}

As is well--known, the maximal Lyapunov exponent $\lambda$ measures the strength chaos in the dynamics of nonlinear systems. Thus, we proceed to systematically calculate the maximal Lyapunov exponent $\lambda $ for our KG and GF models, as well as for 
their corresponding long--range cases, when the energy per particle $\varepsilon $ is gradually increased. We remark that an interesting behavior of the maximal Lyapunov exponent is observed for all cases, as illustrated here in Fig.\ref{KGlyap} for $N=2048$ particles with strong couplings, i.e. $C=1$, which is analyzed and discussed below.

In the zero energy limit,  the classic KG model and the KG--LRI exhibit a power--law dependence 
on the energy $\varepsilon$, namely $\lambda_{KG} \sim \varepsilon ^{1.6}$ and $\lambda^{LRI}_{KG} \sim \varepsilon ^{0.9}$, respectively (see Fig.\ref{KGlyap}(a)). Instead, for the models GF and GF--LRI, we find that they constantly
increase in terms of the energy per particle like $\lambda_{GF}, \lambda^{LRI}_{GF} \sim \varepsilon ^{0.9}$ 
(see  Fig.\ref{KGlyap}(b)), similarly to the KG--LRI system.

As we now progressively increase the energy, the maximal Lyapunov exponent reaches its maximum value at a fixed energy value 
$\varepsilon = \varepsilon_{cr}$ in both  KG systems, 
while it decreases again as $\varepsilon \rightarrow \infty $, according to Fig.\ref{KGlyap}(a). However, for the GF systems 
we find that $\lambda_{GF}, \lambda^{LRI}_{GF} \sim \varepsilon ^{0.27}$ (Fig.\ref{KGlyap}(b)), 
which means the maximal Lyapunov 
exponent is not reaching a maximum as is the KG case, but continues increasing with $\varepsilon $.


Let us try to analyze this behavior, denoting by $V_2=\sum \frac{1}{2} x_n^2$  the quadratic
and by $V_4=\sum \frac{1}{4} x_n^4$ the quartic on-site potentials. For the KG systems, we argue that 
$V_2+W$ represents all quadratic terms and $V_4$ the quartic  ones, while for the GF systems,
the potential function is decomposed into $V_2$ and $V_4+W$, the latter of which contains the quartic terms.
As expected, in all of four systems, when $\varepsilon <<1$ the quadratic terms prevail over the nonlinear ones,
while the opposite happens in the high energy limit $\varepsilon >>1$, where nearly the total potential energy is concentrated in the quartic terms.
However, for the KG and KG--LRI systems this indicates that they tend 
towards an integrable system of {\it uncoupled nonlinear oscillators}: 
$$H_0=\sum_{n}  \frac{1}{2} p_n^2  + \frac{1}{4} x_n^4 $$ 
acting as $N$ self--driven particles. It is therefore expected that above a critical energy 
$\varepsilon_{cr}$ the maximal Lyapunov exponents will decrease (inset of Fig.\ref{KGlyap}(a)).

As a final remark, the GF and GF--LRI models have a similar behavior in terms of their exponents, 
at all energy ranges. It is worth mentioning that the behavior of the maximal Lyapunov exponent 
in the GF and GF--LRI models for $\varepsilon >>1$ is reminiscent of the Fermi--Pasta--Ulam--$\beta$ 
model, where $\lambda  \sim  \varepsilon ^{1/4}$  \cite{casetti}, and which includes both types of 
coupling terms, namely $1/2(x_{n+1}-x_n)^2$ and $1/4(x_{n+1}-x_n)^4$.

\subsection{Approaching an ``integrable'' behavior in the thermodynamic limit \label{gc2}} 

\begin{figure*}
\centering
\resizebox{0.75\columnwidth}{!}{
\includegraphics{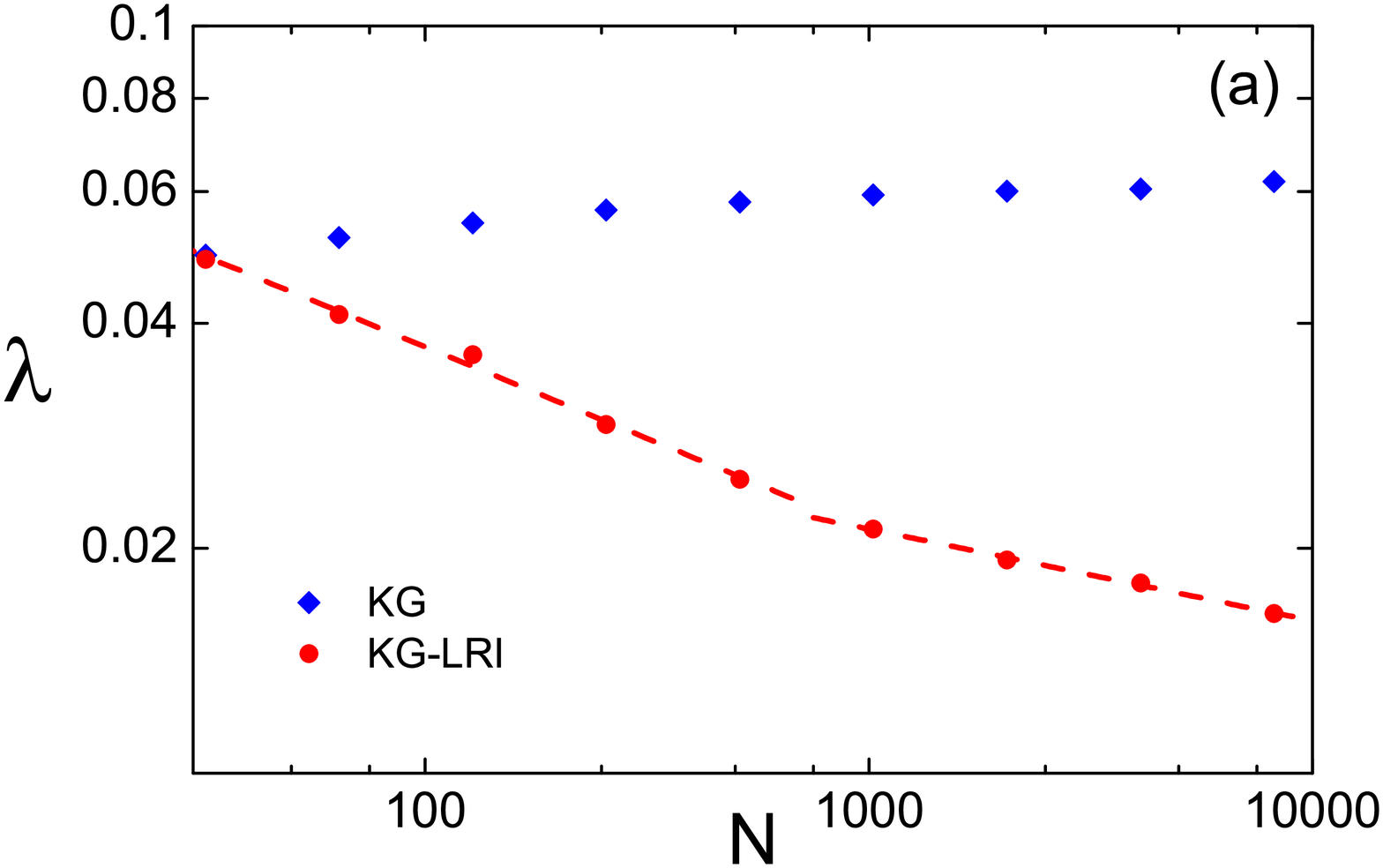}
\includegraphics{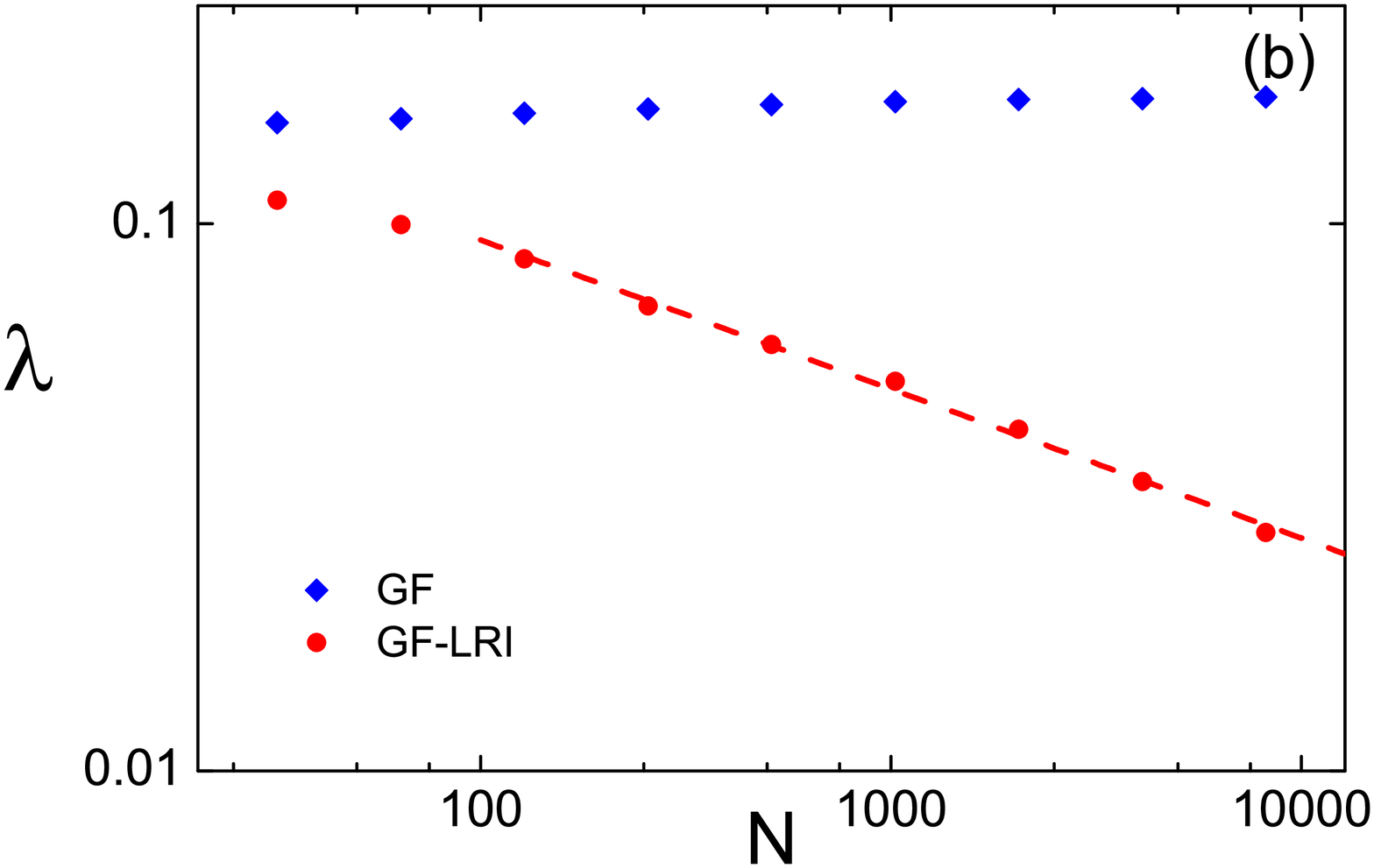}}
\caption{ The maximal Lyapunov exponent for an increasing number of particles $N$. 
(a) KG (blue) diamonds and KG--LRI (red) circles.
(b) GF (blue) diamonds and GF--LRI (red) circles.
\label{lyapN}  }
\end{figure*}
A surprising result awaits us regarding the behavior of the maximal Lyapunov exponent when the number of particles $N$ grows to larger and larger values. Indeed, when we take $\varepsilon=1$ and compare the behavior of $\lambda$ in the classic and long--range KG and GF models, we find a striking difference between short and long interactions. As Fig.~\ref{lyapN}(a),(b) clearly show, the range of interactions has a dramatic effect on the chaotic properties of both systems: While, under short--range interactions, $\lambda$ tends to saturate to a positive value, under LRI $\lambda$ decreases towards zero, thus indicating that both systems become less and less chaotic as $N$ increases. 

In Fig.\ref{lyapN}(a) there are two best line fits describing the decay of $\lambda^{LRI}_{KG} $ of the KG system: A power-law  $N^{-0.24}$ for the data up to $N\simeq 1000$, which slows down to $N^{-0.12}$ for larger $N$. In Fig.\ref{lyapN}(b), on the other hand, the maximal Lyapunov exponent of the GF model exhibits a clear decay with $N$, that is found to scale as $\lambda^{LRI}_{GF} \sim  N^{-0.27}$. At this point it is worth mentioning that the exponent $0.27$ is closer to $1/4$ than the $1/3$ value found for the HMF model \cite{Anteneodo,latora,CirtoAssisTsallis2013} and the FPU--LRI in the absence of quadratic terms \cite{Bagchi}. 

Both lattices in the large $N$ limit approach integrable--like behavior. 
There is, however, one distinct difference between the two models:  When LRI are imposed to the KG system, the decrease of $\lambda$ slows down for $N>1000$, while it continues to decay steadily in the GF model. The reason for this difference is 
presently not known to us and constitutes one more interesting point that merits further investigation.

\section{Momentum distributions of weak and strong chaos at thermal equilibrium \label{md}} 

\begin{figure*}
\centering
\resizebox{0.75\columnwidth}{!}{
\includegraphics{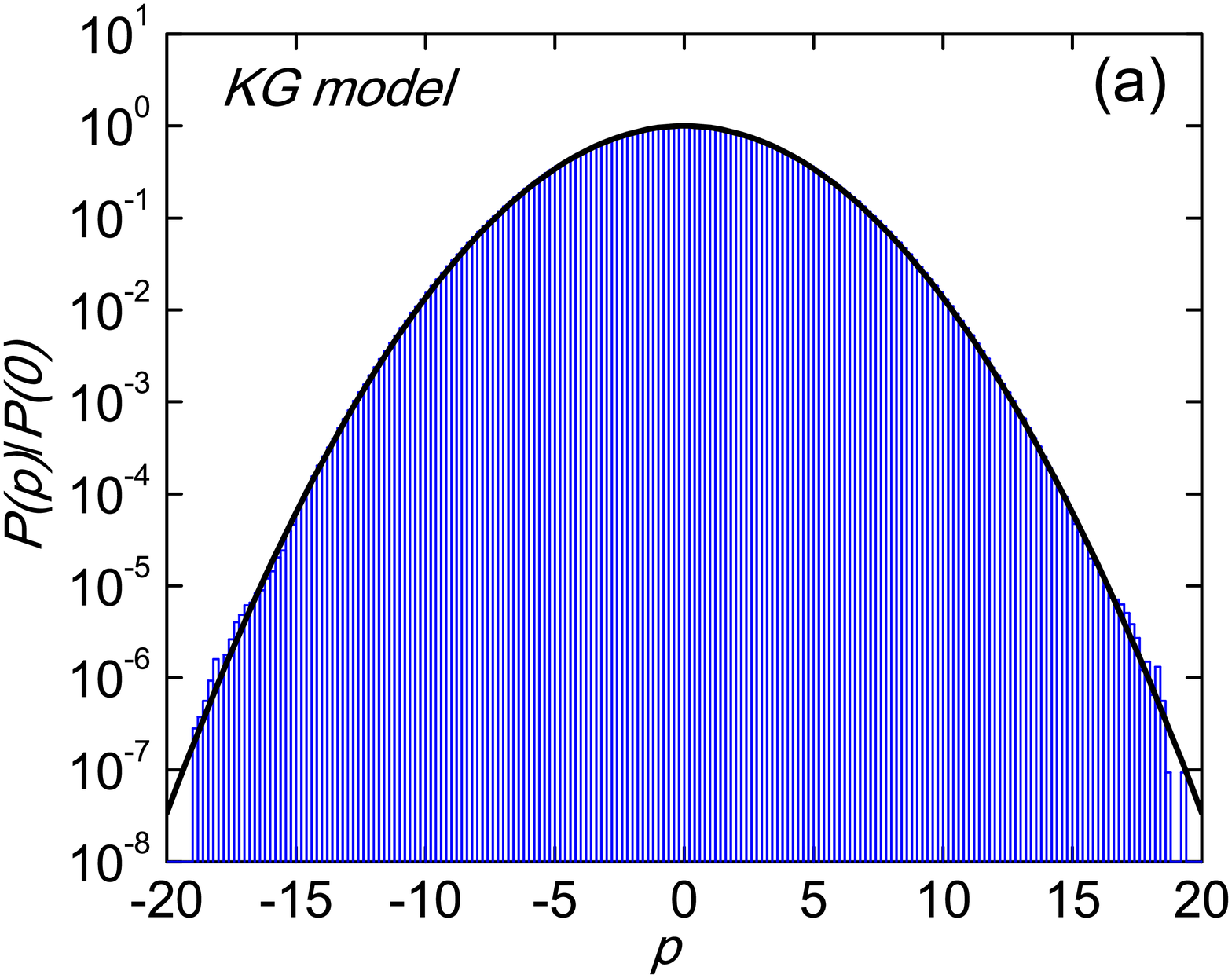}
\includegraphics{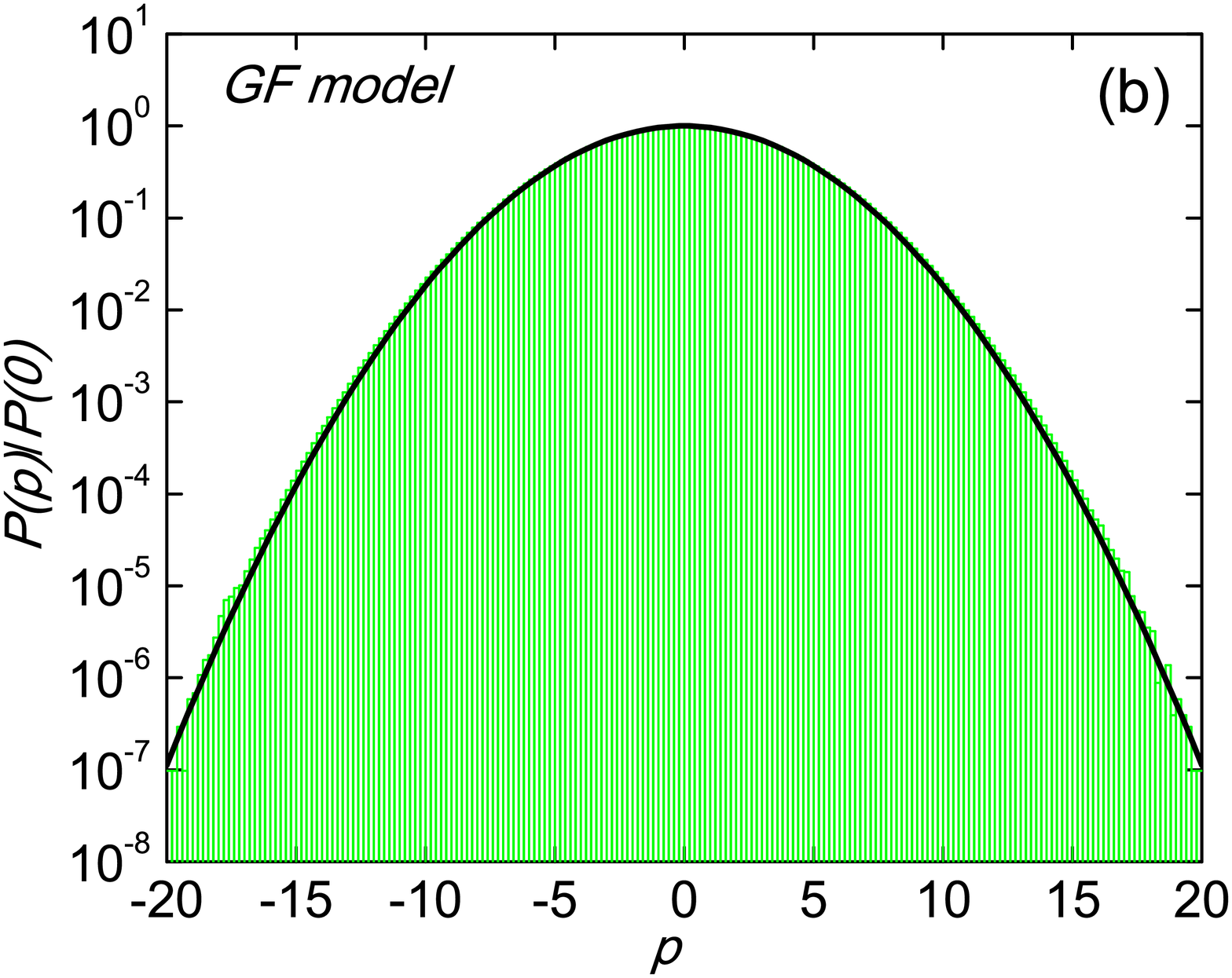}}\\
\resizebox{0.75\columnwidth}{!}{\includegraphics{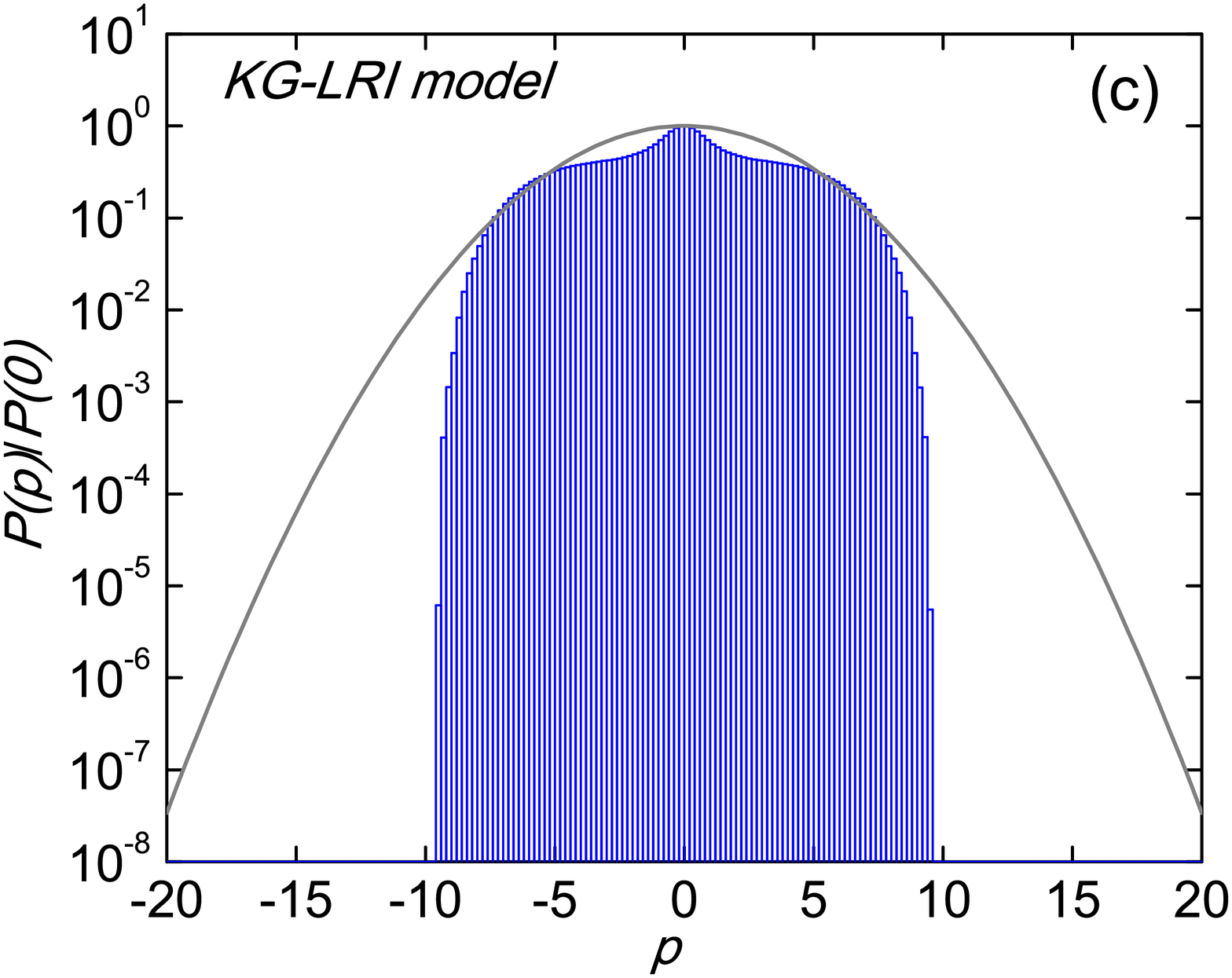}
\includegraphics{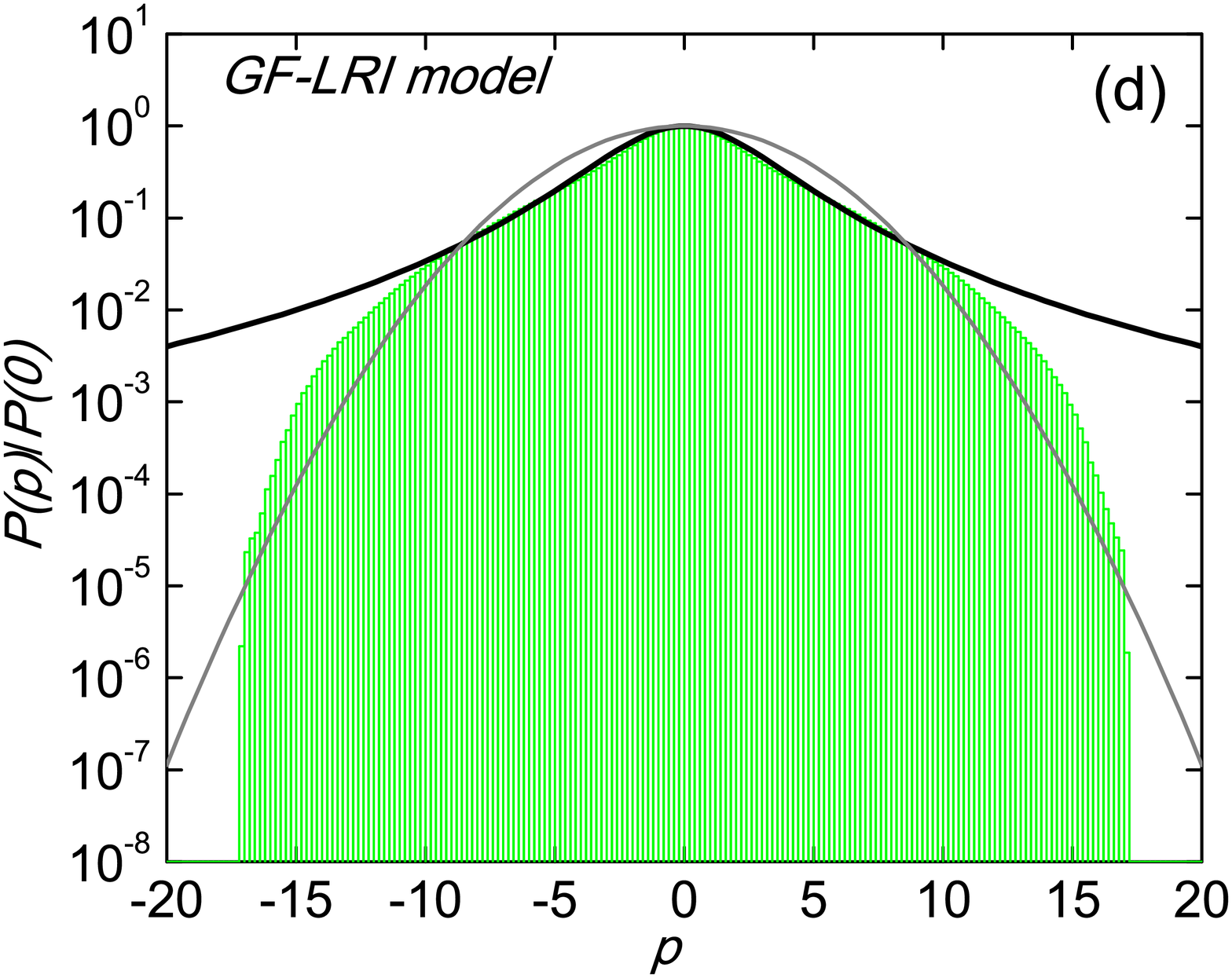}}
\caption{ Time-averaged momentum distributions of 4 cases with $N=2048$ and $\varepsilon =10$ in the
(a) KG, (b) GF, (c) KG--LRI and (d) GF--LRI models respectively. 
\label{dis}  }
\end{figure*}

Finally, let us focus on the statistical properties of the systems studied in this paper. With initial momenta drawn 
randomly from a uniform distribution, as described in Section \ref{gc}, we shall evaluate momenta distributions 
which result from time averages for fixed $N=2048$ and $\varepsilon =10$. 
Our distributions are calculated in the same way as in \cite{EPL,JSTAT}, at times long enough for the pdfs 
to have reached a stationary shape. In particular, we evaluate these time--averaged pdfs at discrete times of an
appropriately chosen step, so as to avoid possible correlations, and we then
assign to the $i$-th momentum band (i.e. $i$-th column of the histograms) the number of times where each of the
momenta $p_1,\ldots,p_N$  falls in, at each of the discrete times.

As we demonstrate in Fig.~\ref{dis}, the resulting pdfs for short and long--range distributions are distinctly different: In the classic (nearest--neighbor) KG and GF models the pdfs quickly approach a Gaussian, suggesting a state of strong chaos, shown in Fig.~\ref{dis}(a),(b). On the other hand, when LRI are imposed, the corresponding distributions are quite far from Gaussians, as Fig.~\ref{dis}(c),(d) clearly show.

Interestingly, even though they are associated with dynamically weakly chaotic regimes and non--equilibrium thermodynamics, 
the KG and GF pdfs under LRI, are not well approximated by $q$--Gaussians, as was the case in our earlier studies on the FPU--$\beta$ lattice \cite{EPL,JSTAT}. This difference may be attributed to the fact that, unlike the FPU, the KG and GF models of the present paper possess on--site potentials, which break translational invariance. This suggests that the thermostatistics of weak chaos in LRI 1D Hamiltonian lattices remains an open subject, that definitely requires further investigation.

\section{Conclusions \label{concl}}
The present study aims to elucidate the role of linear and nonlinear long-range interactions on the dynamical and statistical behavior of systems possessing on-site potentials known for supporting the emergence of discrete breathers. To this end, we have chosen two models, the Klein--Gordon lattice which includes a non-trivial frequency band and the Gorbach--Flach model, which contains only non-linear coupling terms. Simple experiments of exciting only one particle show that the energy spreads  in the KG model, while it remains to a large extent localized in the GF model. Still, in both cases, the presence of long-range interactions causes a dramatic change in the system's behavior, as the initially excited particle preserves its energy to a large degree for as long as we have integrated the equations. One can therefore conclude that such LRI--systems are excellent candidates for studying a variety of different kinds of localized solutions.

When more complicated initial conditions are imposed, such as random momenta excitations of all the particles in the KG--LRI and GF--LRI models, there is an overall dynamical ``regularization' towards the thermodynamic limit.
This regularization is quantified by a power--law decay $N^{-\mu }$, $\mu >0$ of the maximal Lyapunov exponent $\lambda $, 
which indicates a weaker form of chaos and a possible integrable--like behavior as $N \rightarrow \infty $. 
These results are reinforced by a non--trivial statistical behavior of the corresponding momentum distributions, which
however, do not fit a precise $q$--Gaussian shaped form. To this end, we believe that our findings open a variety of interesting questions, which are deferred to a future study. Notwithstanding the vast unexploited properties of systems whose components interact via long--range forces, the above results suggest a remarkably organized behavior, which becomes more and more regular as the degree of complexity and the number of particles increase.
 
\section{Acknowledgements }
The authors acknowledge interesting discussions with Professor C. Tsallis on the results of this paper. A.B. expresses his gratitude to Nazarbayev University for an ORAU grant 2017-2020 supporting his research, and a Social Policy grant that supported his participation to a Dynamics Days Conference at Puebla, Mexico, in November 2016, where some of the results reported in this paper were presented. H.C. was supported by the State Scholarship Foundation (IKY) operational Program: `Education and Lifelong Learning--Supporting Postdoctoral Researchers' 2014-2020, and is co--financed by the European Union and Greek national funds.
%

\end{document}